\documentclass[preprint,3p,twocolumn]{elsarticle}
\usepackage{graphicx,color,bm,mathtools,amsfonts}
\usepackage[utf8]{inputenc}
\usepackage{hyperref}

\def\d{\text{d}}

\newcommand{\bra}[1]{\left\langle #1 \right|}
\newcommand{\ket}[1]{\left|#1\right\rangle}
\newcommand{\braket}[2]{\left\langle #1 \right|\left.\!\! #2 \right\rangle}
\newcommand{\mr}[1]{\mathrm{#1}}
\newcommand{\qbar}{\bar{q}}
\def\a{\alpha} \def\b{\beta}
\newcommand{\as}{\alpha_s}
\newcommand{\asb}{\bar{\alpha}_s}

\def\X{{\scriptscriptstyle X}}
\def\R{{\scriptscriptstyle\mathrm{R}}}
\def\V{{\scriptscriptstyle\mathrm{V}}}

\newcommand{\overbar}[1]{\mskip.9\thinmuskip\overline{\mskip-.9\thinmuskip {#1} \mskip-.9\thinmuskip}\mskip.9\thinmuskip}

\def\cA{\mathcal{A}}  \def\cAb{\bar{\mathcal{A}}}
\def\cB{\mathcal{B}}  \def\cBb{\bar{\mathcal{B}}}

\def\cC{\mathcal{C}}

\def\cG{\mathcal{G}}

 \def\cNb{\overbar{\mathcal{N}}}
\def\cQ{\mathcal{Q}}
\def\cW{\mathcal{W}}
\def\cWI{\overline{\mathcal{W}}}
\def\cWIR{\widetilde{\overline{\mathcal{W}}}}
\def\cWII{\overline{\overline{\mathcal{W}}}}
\def\cWR{\widetilde{\mathcal{W}}}

\def\bA{\mathbf{A}}  \def\bAb{\bar{\mathbf{A}}}

\def\bT{\mathbf{T}}

\def\fA{\mathfrak{A}}

\def\CF{\mathrm{C_F}}

\def\CA{\mathrm{C_A}}
\def\CAsq{\mathrm{C_A^2}}
\def\CAcub{\mathrm{C_A^3}}

\begin{document}

\begin{frontmatter}

\title{Eikonal amplitudes for three-hard legs processes at finite-N$_c$}

\author[KKaddress]{K. Khelifa-Kerfa\corref{CorresAuthor}}
\address[KKaddress]{Department of Physics, Faculty of Science \\ Islamic University of Madinah, Madinah 42351, Saudi Arabia}
\cortext[CorresAuthor]{Corresponding author}
\ead{kamel.kkhelifa@gmail.com}   

\author[YDaddress]{Y. Delenda}
\address[YDaddress]{Laboratoire de Physique des Rayonnements et de leurs Int\'{e}ractions avec la Mati\`{e}re, \\
D\'{e}partement de Physique, Facult\'{e} des Sciences de la Mati\`{e}re,\\
Universit\'{e} de Batna--1, Batna, Algeria}

\begin{abstract}
We extend our previous work on scattering amplitudes \cite{Delenda:2015tbo} to hadron collisions. We provide a general formalism for the computation of eikonal amplitudes squared for the radiation of soft and energy-ordered gluons off three massless hard-legs at finite-N$_c$ to any order in the perturbative expansion. 
Examples of three-hard legs processes include vector/Higgs boson production in association with a single hard jet and dijet production in DIS.
Explicit expressions for the radiation of up to four gluons are provided as an illustration of the formalism.    
\end{abstract}


\end{frontmatter}


\section{Introduction}
\label{sec:Intro}
As the Large Hadron Collider (LHC) continues to produce high-precision experimental data the theory community is required to devote more efforts on the corresponding theoretical calculations in order to match and/or surpass their experimental counterpart results. Not only will such high-precision calculations shed light on some of the unresolved issues of the current Standard Model, but will contribute to the endeavour for new physics discoveries. 

Remarkable theoretical advances have been achieved during the last two decades, thanks to the development of new  techniques both analytically and numerically along with the help of powerful computers. For instance, processes such as Higgs production have been computed at fixed-order (FO), in the QCD strong coupling $\as$, up to next-to-next-to-next leading order (N$^3$LO) \cite{Anastasiou_2014, Anastasiou_2015}  and to  all-orders up to next-to-next-to-next leading logarithms (N$^3$LL) for related  differential distributions \cite{Bizo_2018, Chen_2019}. Other processes with similar FO accuracy include the determination of evolution kernels of parton distribution functions (PDFs) \cite{Moch_2017, Moch_2018}, IR structure of multileg QCD scattering amplitudes \cite{Almelid_2016}, beta function and cusp anomalous dimension \cite{Harlander_2006, henn2019fourloop, von_Manteuffel_2020}, to name few. Such computations are, however, limited to  few processes and involve substantial efforts and numerous complex computer codes.   

A more universal and process-independent feature of QCD scattering amplitudes is their form at soft and/or collinear regions of phase space. The latter form captures the most singular part of the amplitude and hence represents the leading dominant contribution at those regions. Moreover, soft/collinear singularities are known to factorise at all-orders in perturbation theory (PT), exhibit a pattern of symmetry and may thus be computed to all-orders with much less technical effort compared to their counterpart exact formulae. This has paved the way for a better understanding of the structure of QCD radiation amplitudes and has been used for the resummation of large logarithms, that arise at the said soft/collinear kinematical regions due to mis-cancellation between real emissions and virtual corrections, up to N$^{3}$LL accuracy \cite{Bizo_2018, Chen_2019}. It has also been exploited in various manners in the determination of relevant parts of many exact FO results (see, for instance, \cite{bendavid2018les} and references therein). Furthermore, Monte Carlo event generators, such as \texttt{Pythia}, \texttt{Herwig} and \texttt{Sherpa} rely on the universal properties of soft/collinear singularities in their parton showers' algorithms (see \cite{Buckley_2011} for a review). 

In our previous work \cite{Delenda:2015tbo} we presented a general formula for the computation of eikonal (soft) amplitudes squared for the radiation of arbitrary soft, energy-ordered gluons in the $e^+ e^-$ annihilation process: $e^+ e^- \to q \qbar + g_1 + \cdots + g_n$. In the present paper we  extend the latter work to hadron collisions involving up to three massless hard-legs which include, for instance, vector and Higgs bosons production in association with a hard jet as well as dijet production in Deep Inelastic Scattering (DIS). The computation of amplitudes and squared amplitudes is automated with the help of the \texttt{Mathematica} program of \cite{Delenda:2015tbo} after the appropriate modifications are implemented. 

It is worth noting that the issue of soft/eikonal gluon radiation amplitudes beyond two-loops has been addressed in \cite{Catani_2020} for two- and three-hard legs both for massless and massive partons, with and without strong-energy ordering. Our work herein, which is based on our previous work  \cite{Delenda:2015tbo} that appeared before \cite{Catani_2020}, serves as an independent cross check for the findings of \cite{Catani_2020} for the specific case of massless partons in the strong-energy ordering regime. Moreover, we present explicit expressions for the eikonal amplitudes squared for the emission of up to four soft gluons off three-hard legs. In \cite{Catani_2020} emission amplitudes were explicitly given only for up to three soft gluons.  

This letter is organised as follows. In section \ref{sec:Two-hard legs} we review our work in \cite{Delenda:2015tbo} and extend it to the case of arbitrary two-hard legs involving both quarks and gluons. An example of a process involving two-hard gluons is the Higgs production via gluon fusion \cite{Ellis:1987xu}.  We then present, in section \ref{sec:Three-hard legs}, the formalism of eikonal amplitudes squared for the emission of soft energy-ordered gluons off three-hard legs at hadron collisions. In the same section we present the explicit expressions of the said eikonal amplitudes squared up to four-loops. We conclude in section \ref{sec:Conclusion}.

\section{Two-hard legs: revisited}
\label{sec:Two-hard legs}

In this section we review our previous work \cite{Delenda:2015tbo} and introduce the necessary notations and concepts that we need for later development. The interested reader is advised to consult the latter reference for a concrete understanding of the current letter.   

The (partonic) differential cross-section for the emission of $m$ soft energy-ordered (real) gluons for a two-hard (massless) legs process such as $e^+ e^- \to p_a + p_b + g_1 + \cdots + g_m$ is given by \cite{Delenda:2015tbo}:
\begin{align}\label{eq:Diff-XSec_general}
 \d \sigma_m 
 = \d\Phi_m \bra{0} W_m \ket{0},
\end{align}    
where $\ket{0}$ is the Born amplitude, which is a vector in colour space and is proportional to $\delta_{c_a c_b} \ket{c_a, c_b}$ with $c_a, c_b$ being the colour indices of the two hard partons $p_a$ and $p_b$, respectively. For quarks (and anti-quarks) these indices take the values $1, \dots, N_c$, while for gluons they run from $1, \dots, N_c^2-1$, where $N_c$ is the number of colour charges in the fundamental representation (degree of SU$(N_c)$ group). The phase space factor, $\d\Phi_m$, is given by:  
\begin{align}\label{eq:PhaseSpaceFactor_general}
\d\Phi_m = \prod_{i=1}^{m} \frac{k_{ti} \d k_{ti}}{4 \pi^2}\, \frac{\d\eta_i \d\phi_i}{4 \pi},
\end{align} 
where $k_{ti}$ is the transverse momentum of the $i^{\mr{th}}$ emitted gluon, $\phi_i$ its azimuthal angle and $\eta_i$ its rapidity. The emission amplitude squared, $W_m$, is given by \cite{Delenda:2015tbo}: 
\begin{align}\label{eq:EmissionAmpSq_general}
 W_m = (-1)^m\, g_s^{2m}\, \left( \prod_{n=1}^m \sum_{i_n \neq j_n \in U_{n-1}}  \omega_{i_n j_n}(k_n)  \right) \times 
 \notag \\ 
\times   (\bT_{j_1}^{a_1})^\dagger \cdots (\bT_{j_m}^{a_m})^\dagger \; \bT_{i_m}^{a_m} \cdots \bT_{i_1}^{a_1}, 
\end{align}
where $g_s$ is the strong coupling and $\omega_{ij}(k)$ is the dipole antenna function defined by: 
\begin{align}\label{eq:DipoleAntenna_definition}
 \omega_{ij}(k) = \frac{(h_i \cdot h_j)}{(h_i \cdot k)(k \cdot h_j)},
\end{align}
with $h_\ell$ and $k$ being the four-momenta of the emitting parton $\ell$ and the emitted gluon, respectively. The set $U_{n-1} = \{a,b, 1, \cdots, n-1\}$ represents the set of all possible emitters of the softest gluon $n$. 
The colour operators $\bT_i^{c}$ act on the colour space of parton $i$ by inserting a gluon with colour index $c$ and repainting parton $i$. That is, for an initial state with, say, two partons the action of the colour operator reads:    
\begin{align}
 \bT_2^c \ket{c_1, c_2} = T_{c'_2 c_2}^c \ket{c_1, c'_2, c}.
\end{align}
The adjoint operator $(\bT_i^c)^\dagger$ plays the opposite role to that of $\bT_i^c$. More details of the properties of these colour operators are reported in \cite{Delenda:2015tbo}. Here we note that $(\bT_i^c)^2 = C_i$ the Casimir scalar which equals to $\CF = (N_c^2-1)/2N_c$ if parton $i$ is a quark (anti-quark) and $\CA = N_c$ if parton $i$ is a gluon. We also recall the identity (see eq. (21) of \cite{Delenda:2015tbo}): 
\begin{align}\label{eq:ColorOperator_Identity}
 \sum_{i \in U_{m-1}} \bT_i^{a_m} |m-1\rangle = 0.
\end{align} 
This shows that the $|m+1 \rangle$ ($m-1$ gluons + two-hard legs) transforms as a colour singlet under SU(3). Note that the generators $\bT_i^a$ are taken as if all partons were {\it incoming}. 

Substituting Eqs. \eqref{eq:PhaseSpaceFactor_general} and \eqref{eq:EmissionAmpSq_general} into the formula of the differential cross-section \eqref{eq:Diff-XSec_general} we can recast the latter in the form: 
\begin{align}\label{eq:Diff-XSec_general2}
\d \sigma_m &= \braket{0}{0} \sum_{m=1}^\infty \, \asb^m \left(\prod_{i=1}^{m} \frac{\d k_{ti}}{k_{ti}} \d\eta_i \frac{\d\phi_i}{4 \pi}  \right)\, \times
\notag\\
& \hspace{1cm} \times \sum_{\X} \cW^\X_{12\hdots m}, 
\end{align}
where $\asb = \as/\pi = g_s^2/(4\pi^2)$ and the superscript $X$ stands for a given gluon configuration in the real/virtual sense. That is, $X = \mr{\small{x_1 x_2 \hdots x_m}}$ where $\mr{\small x_i} \in \{\mr{R,V} \}$, which means that a given gluon $i$ may either be a real or virtual gluon. The treatment of virtual gluons is analogous to that of real gluons in the eikonal approximation, both for amplitude squared and phase space, as was shown in \cite{Delenda:2015tbo}. The main difference stems from the ordering of the colour operators in Eq. \eqref{eq:EmissionAmpSq_general}.    

The form of the eikonal amplitude squared $\cW^\X_{12 \hdots m}$ for a given gluon configuration $X$ reads: 
\begin{align}\label{eq:EikAmp_2HardLegs_Final}
 \cW^{\X}_{1\hdots m} &= (-1)^m\, \left( \prod_{i=1}^{m} \sum_{i_n \neq j_n \in U_{n-1}} w_{i_n j_n}^n \right)  \cC^{i_1 i_2 \hdots i_m}_{j_1 j_2 \hdots j_m} ,
\end{align}
where $w_{i_n j_n}^n = k_{tn}^2 \, \omega_{i_n j_n}(k_n)$ and $\omega_{ij}(k)$ is defined in \eqref{eq:DipoleAntenna_definition}, i.e., $w_{ij}^n$ is purely an angular function, and the colour factor reads: 
\begin{align}\label{eq:ColourFactor_2HardLegs_final}
\cC^{i_1 i_2 \hdots i_m}_{j_1 j_2 \hdots j_m} &= \frac{1}{ \braket{0}{0} } \bra{0} (\bT_{j_1}^{a_1})^\dagger \cdots (\bT_{j_m}^{a_m})^\dagger \;
\notag\\
& \hspace{2cm} \bT_{i_m}^{a_m} \cdots \bT_{i_1}^{a_1} \ket{0}.
\end{align} 
Notice that in Eq. \eqref{eq:Diff-XSec_general2} the Born amplitude squared has been factorised out. 
A list of properties of the eikonal amplitude squared \eqref{eq:EikAmp_2HardLegs_Final} including its symmetries with respect to the various partons, its colour structure at higher loops, a description of the \texttt{Mathematica} program used to compute it and its explicit formulae for up to five-loops were given in \cite{Delenda:2015tbo}. Although the latter formulae were given for the specific case: $a=q$ and $b=\qbar$, the formulae \eqref{eq:EikAmp_2HardLegs_Final} and \eqref{eq:ColourFactor_2HardLegs_final} apply to any pair of partons, including both quarks/anti-quarks and gluons.  

In the next section we present a general form for the eikonal amplitude squared for the emission of $m$ soft energy-ordered gluons for processes with three-hard legs.

\section{Three-hard legs}
\label{sec:Three-hard legs}

Examples of processes that involve three-hard legs include dijet production in DIS, vector ($\gamma, Z, W$) and Higgs ($H$) boson production in association with a single jet at hadron colliders. Here we shall focus on the latter but the results we report are equally valid for the former and for any massless three-hard legs processes in general. The aforementioned bosons are colour neutral and thus play no role in our calculations. There is a total of four partonic channels that contribute to the three-hard legs Born process: $p_a + p_b \to p_c + B$, where $B$ stands for one of the bosons $\gamma, Z, W$ or $H$. Three of which are for vector boson production + jet, which read: 
$(\delta_1) = \{q\qbar g\}$: $q\qbar \to g+B$, 
$(\delta_2) = \{qgq\}$: $q g \to q+B$ and 
$(\delta'_2) = \{\qbar g \qbar\}$: $\qbar g\to \qbar +B$, and the last channel is for the Higgs production + jet, which reads: 
$(\delta_3) = \{ggg\}$:  $g g \to g+H$. Notice that for our calculations channels ($\delta_2$) and ($\delta'_2$) are identical and thus only the former will be considered. 

The emission of $m$ soft energy-ordered gluons off the said three-hard legs Born channels is schematically represented as: 
\begin{align}
 p_a + p_b \to p_c + B + g_1 + \cdots + g_m.
\end{align}  
Unlike the two-hard legs case, where the hardest gluon $g_1$ may only be emitted by a single dipole, namely $(ab)$, here $g_1$ may be emitted by any of the three possible dipoles $(ab), (ac)$ and $(bc)$. In this sense it is analogous to the emission of $g_2$ for the two-hard legs case as the latter may also be emitted by three dipoles, namely $(ag_1), (bg_1)$ and $(ab)$. Hence it is possible to exploit this analogy in order to derive three-hard legs emission amplitudes squared from those of two-hard legs. We shall employ both this analogy and the \texttt{EikAmp} program, as an independent confirmation, to derive the form of the eikonal amplitudes for three-hard legs. 

The general form of the eikonal amplitude for the emission of $m$ soft energy-ordered gluons off a given  three-hard legs Born channel $(\delta)$ and for a particular gluon configuration $X$ is given by an expression identical to \eqref{eq:EikAmp_2HardLegs_Final} with two modifications: (i) The set of possible emitters becomes $U_{n-1} = \{a,b,c, 1, \hdots, n-1\}$, and (ii) the Born amplitude $\ket{0}$ (and the associated colour structure) is different from that of two-hard legs and is different for each channel $\delta$.
The treatment of virtual gluons is identical to that of two-hard legs, which was discussed in details in Ref. \cite{Delenda:2015tbo}. 

In the next subsections we present explicit formulae for the eikonal amplitudes squared up to four-loops for each of the three partonic channels mentioned earlier.

\subsection{One-loop}
\label{subsec:3HardLegs-1loop}

The one-loop eikonal amplitudes squared may be given in the compact form:  
\begin{subequations}\label{eq:EikAmp-1loop}
\begin{align}\label{eq:EikAmp-1loop-R}
 \cW_{1,\delta}^{\R} &= \sum_{(ij) \in \Delta_\delta} \cC_{ij}\, w_{ij}^1,
  \\
 \cW_{1,\delta}^{\V} &= - \cW_{1,\delta}^\R, 
 \label{eq:EikAmp-1loop-V}
\end{align}
\end{subequations}
where the sum is over all possible dipoles in the partonic Born channel $\delta$. That is, given that $\delta = \{abc\}$ then $\Delta_\delta =  \{(ab), (ac), (bc)\}$. The {\it dipole} colour factor $\cC_{ij}$ is defined as: 
\begin{align}\label{eq:DipoleColorFactor}
 \cC_{ij} = - 2\,\bT^a_i \cdot \bT^a_j, 
\end{align}
where the colour operators $\bT^a_{i}$ have been defined and discussed in the previous section and we note that $\bT^a_i \cdot \bT^a_j = \bT^a_j \cdot \bT^a_i$. Using the identity \eqref{eq:ColorOperator_Identity} we have for the three channels: 
\begin{align}\label{eq:DipoleColorFactor_AllChannels}
(\delta_1)&: \qquad \cC_{ab} = 2\CF-\CA, \qquad \cC_{aj} = \cC_{bj} = \CA, 
\notag\\
(\delta_2)&: \qquad\cC_{ab} = \cC_{bj} = \CA, \qquad \cC_{aj} = 2\CF-\CA,
\notag\\
(\delta_3)&: \qquad\cC_{ab} = \cC_{bj} = \cC_{aj} =  \CA.
\end{align}
It is then straightforward to obtain the explicit form of \eqref{eq:EikAmp-1loop} for each channel,  by substituting \eqref{eq:DipoleColorFactor_AllChannels} back into \eqref{eq:EikAmp-1loop}, in terms of the colour Casimir scalars, $\CF$ and $\CA$. Note that Eq. \eqref{eq:EikAmp-1loop-R} may be shown, upon  substituting the expressions in \eqref{eq:DipoleColorFactor_AllChannels}, to be exactly identical to Eq. (6.7) of Ref. \cite{Catani_2020} after taking into account redefinitions of dipole colour factors. 
Our form \eqref{eq:EikAmp-1loop-R} is more compact and manifests the dipole structure of the eikonal amplitudes squared. Moreover, amplitudes squared for other gluon configurations which involve virtual gluons such as \eqref{eq:EikAmp-1loop-V} are not explicitly given in \cite{Catani_2020} in any approximation.  

\subsection{Two-loops}
\label{subsec:3HardLegs-2loop}

The two-loops eikonal amplitudes squared for a given channel $\delta$ for the various gluon configurations are given by analogous forms to those for two-hard legs (see Eq. (41) of \cite{Delenda:2015tbo}): 
\begin{subequations}\label{eq:EikAmp-2loop}
\begin{align}
 \cW_{12, \delta}^{\R\R} &= \cW_{1,\delta}^\R \cW_{2,\delta}^{\R} + \cWI_{12,\delta}^{\R\R}, \qquad \cW_{12,\delta}^{\R\V} = - \cW_{12,\delta}^{\R\R},
\\
\cW_{12,\delta}^{\V\R} &= - \cW_{1,\delta}^\R \cW_{2,\delta}^\R, \qquad \cW_{12,\delta}^{\V\V} = -\cW_{12,\delta}^{\V\R}.   
\end{align}
\end{subequations}
The term $\cW_{1,\delta}^\R \cW_{2,\delta}^\R$, which we shall henceforth refer to as the {\it reducible} contribution, represents successive {\it independent} emissions of gluons, with $\cW_i^\R$ given in Eq. \eqref{eq:EikAmp-1loop-R}. The term $\cWI_{12,\delta}^{\R\R}$, on the other hand, represents a new contribution, which is not related to the one-loop result, and which comes from correlated gluon emissions. We refer to the latter as the {\it irreducible} contribution, and at this order it is given by:
\begin{align}\label{eq:EikAmp-2loop-Irred}
 \cWI_{12,\delta}^{\R\R} = \CA  \sum_{(ij) \in \Delta_\delta} \cC_{ij}\, \cA^{12}_{ij}, 
\end{align}
where we have introduced the two-loop antenna function
\begin{align}\label{eq:AntennaFunc-2loop}
 \cA_{\a \b}^{ij} = w_{\a\b}^i (w_{\a i}^j + w_{i\b}^j - w_{\a\b}^j).
\end{align}
Once again, substituting the values of the colour factors \eqref{eq:DipoleColorFactor_AllChannels} into Eq. \eqref{eq:EikAmp-2loop-Irred} and simplifying the all-real amplitude squared $\cW_{12,\delta}^{\R\R}$ in \eqref{eq:EikAmp-2loop} one can easily show that the latter is identical to Eq. (6.8) of \cite{Catani_2020} (again after taking into account redefinitions of colour factors). 

We notice that for the one- and two-loop amplitudes squared the corresponding expression for channel  ($\delta_3$) may easily be deduced from that of channel $(\delta_1)$ or ($\delta_2$) by the simple replacement $\CF \to \CA$. Such a replacement is dubbed {\it Casimir scaling} in \cite{Catani_2020}. The latter is broken for the emission of three soft gluons and more, as we shall discuss below \footnote{For the case of two-hard legs, Casimir scaling works fine for one-, two- and three-loops, but breaks at four-loops and beyond.}.

\subsection{Three-loops}
\label{subsec:3HardLegs-3loop}

The three-loops eikonal amplitudes squared for the emission of three soft energy-ordered gluons off  three-hard  legs Born process $\delta$ are given by analogous forms to those of the two-hard legs case (Eqs. (43) in \cite{Delenda:2015tbo}). In particular 
\begin{subequations}\label{eq:EikAmp-3loop}
\begin{align}
 \cW_{123,\delta}^{\R\R\R} &= \prod_{i=1}^3 \cW_{i,\delta}^\R + \prod_{ijk=1}^{3} \cW_{i,\delta}^\R \, \cWI_{jk,\delta}^{\R\R} + \cWI_{123,\delta}^{\R\R\R}, 
 \label{eq:EikAmp-3loop-RRR}
 \\
 \cW_{123,\delta}^{\R\V\R} &= -\prod_{i=1}^3 \cW_{i,\delta}^\R - \prod_{ik=2}^{3} \cW_{i,\delta}^\R \, \cWI_{1k,\delta}^{\R\R} + \cWI_{123,\delta}^{\R\V\R},
 \label{eq:EikAmp-3loop-RVR}
\end{align}
and the other remaining amplitudes squared are analogous to Eqs. (43c), (43d) and (44) of \cite{Delenda:2015tbo} and involve no new contributions. 
The all-real three-loops irreducible contribution $\cWI_{123,\delta}^{\R\R\R}$ is given by the sum:
\begin{align}\label{eq:EikAmp-3loop-I}
 \cWI_{123,\delta}^{\R\R\R} =  \cWIR_{123,\delta}^{\R\R\R} +  \cWII_{123,\delta}^{\R\R\R},
\end{align}
where 
\begin{align}
 \cWIR_{123,\delta}^{\R\R\R} &= \CAsq \sum_{(ij) \in \Delta_\delta} \cC_{ij} \left[\cA_{ij}^{12} \cAb_{ij}^{13} + \cB_{ij}^{123} \right] , 
 \label{eq:EikAmp-3loop-IR}
\\
\cWII_{123,\delta}^{\R\R\R} &= \sum_{\pi_{\{ijk\}} } \, \cQ_{\delta} \left[\cG_{ij}^{k1}(2,3) + 2 \leftrightarrow 3 \right].
\label{eq:EikAmp-3loop-II}
\end{align}
\end{subequations}
The term $\cWI_{123,\delta}^{\R\V\R}$ in \eqref{eq:EikAmp-3loop-RVR} is given by an identical expression to \eqref{eq:EikAmp-3loop-I}, with an overall minus sign, except the term $\cB_{ij}^{123}$ which is absent as gluon $2$ is virtual and cannot thus be an emitter. We have $\cAb_{ij}^{k\ell} = \cA_{ij}^{k\ell}/w_{ij}^k$ and the three-loop antenna function $\cB_{ij}^{k\ell m}$ is defined as: 
\begin{align}\label{eq:AntennaFunc-3loop}
 \cB_{ij}^{k\ell m} = w_{ij}^k \left(\cA_{ik}^{\ell m} + \cA_{kj}^{\ell m} - \cA_{ij}^{\ell m} \right).
\end{align}
Unlike \eqref{eq:EikAmp-3loop-IR}, the term $\cWII_{123,\delta}^{\R\R\R}$ in Eq.  \eqref{eq:EikAmp-3loop-II} is the first contribution which does not resemble any structure seen at previous orders. It is referred to as the ``{\it quadrupole}" contribution in \cite{Catani_2020}. 
It was given in Eqs. (54)-(56) of \cite{Delenda:2015tbo} for the two-hard legs ($q\qbar$) case. Here we have written it down in the following form:
\begin{align}\label{eq:quadrupoleFun}
 \bAb^{k\ell m n }_{ij} = -  \sum_{\pi_{\{ijk\}} } \cG_{ij}^{k \ell}(m,n) + (m \leftrightarrow n), 
\end{align}
where the sum is over the permutation $\pi_{\{ijk\}} = \{(ijk), (ikj), (jki) \}$ and the expression above is the same for all channels, just like all other angular functions defined thus far. In other words, the differences between the various channels are all confined in the colour factor part of the eikonal amplitudes. The angular part of the latter amplitudes is channel-independent.  
The new pseudo-antenna function $\cG_{ij}^{k\ell}(n,m)$ in Eq. \eqref{eq:quadrupoleFun}  may be cast in various ways:
\begin{align}
 \cG_{ij}^{k\ell}(n,m) &= w_{ij}^\ell \, T_{ij}^{k\ell}(n) U_{ij}^{k\ell}(m),
  \notag\\
 &=  w_{ij}^\ell \, T_{ij}^{k\ell}(n) T_{ji}^{k\ell}(m)
 \notag\\
 &= w_{ij}^\ell \, U_{ji}^{k\ell}(n) U_{ij}^{k\ell}(m),
\end{align} 
with the cross ($t$- and $u$-) channel functions defined by: 
\begin{subequations}\label{eq:XChannel-Funs}
\begin{align}
 T_{ij}^{k\ell}(n) &= w_{ij}^n + w_{k\ell}^n - w_{ik}^n - w_{j\ell}^n, 
\\
 U_{ij}^{k\ell}(n) &= w_{ij}^n + w_{k\ell}^n - w_{i\ell}^n - w_{jk}^n. 
\end{align}
\end{subequations}
We note the following symmetry properties for the cross-channel functions:
\begin{align}
T_{ij}^{k\ell} = - T_{\ell j}^{k i} = -T_{ik}^{j\ell}, & \quad 
U_{ij}^{k\ell} = - U_{kj}^{i \ell} = - U_{i\ell}^{kj}
\notag\\
T_{ij}^{k\ell} = U_{ji}^{k\ell} = U_{ij}^{\ell k}, & \quad
T_{ij}^{k\ell} = T_{kj}^{i\ell} + U_{ij}^{k\ell}.
\end{align} 
Various peculiar features have been demonstrated for $ \bAb^{k\ell m n }_{ij}$ in \cite{Delenda:2015tbo} including breaking of {\it Bose} and {\it mirror} symmetries \footnote{See Ref. \cite{Delenda:2015tbo} for full details.}. The colour factor $\cQ_\delta$ in \eqref{eq:EikAmp-3loop-II} is given for each channel by:
\begin{align}\label{eq:Quadrupole-Channels}
 &\qquad \cQ_{\delta_1} = \cQ_{\delta_2} = \CAsq (\CA-2\CF) = \CA, 
\notag\\
& \qquad \cQ_{\delta_3} = 6\, \CA. 
\end{align}
These agree with those reported in \cite{Catani_2020} (Eqs. (B.16) and (B.17)) after redefinitions of colour factors are accounted for. This indicates that our colour factor $\cQ_\delta$ is equivalent to the operator $Q_{abab}$ of \cite{Catani_2020}, where $(ab) = (q\qbar)$ for $\delta_1$ and $\delta_2$, and $(ab)=(gg)$ for $\delta_3$, which is the only quadrupole operator that needs to be explicitly calculated for the three-hard legs case.

\subsection{Four-loops}
\label{subsec:3HardLegs-4loop}

The eikonal amplitudes squared are again given by analogous formulae to those reported in \cite{Delenda:2015tbo}. That is, for a given channel $\delta$ and a given gluon configuration $X$ we write them as a sum of reducible and irreducible contributions: 
\begin{align}\label{eq:EikAmp-4loop}
 \cW_{1234,\delta}^{\X} = \cWR_{1234, \delta}^\X + \cWI_{1234, \delta}^{\X}.
\end{align}
Employing the following property of eikonal amplitudes squared: $\cW_{1\hdots n}^{\mr{x_1} \hdots \V} = - \cW_{1\hdots n}^{\mr{x_1} \hdots \R}$ where $\mr{x_i}$ may either be real (R) or virtual (V), we are left with 8 out of 16 amplitudes squared to determine. They may actually all be deduced from the all-real RRRR amplitude squared as explained in Sec. C.1 of \cite{Delenda:2015tbo}. 
The reducible and irreducible parts of the latter all-real amplitude squared are given by:
\begin{subequations}
\begin{align}\label{eq:EikAmp-4loop-R}
 \cWR_{1234, \delta}^{\R\R\R\R} &= \prod_{i=1}^4 \cW_{i,\delta}^\R + \sum_{ijk\ell=1}^4 \big[ \cW_{i,\delta}^\R \cW_{j,\delta}^\R \cWI_{k\ell,\delta}^{\R\R} \Theta^{i<j}
 \notag\\
&+ \cW_{i,\delta}^\R \cWI_{jk\ell, \delta}^{\R\R\R} + \cWI_{ij,\delta}^{\R\R} \cWI_{k\ell,\delta}^{\R\R} \Theta^{i < k}  \Big], 
\end{align}
where $\Theta^{i<j} = \Theta(j-i)$ is the heaviside theta function, and 
\begin{align}\label{eq:EikAmp-4loop-I}
 \cWI_{1234,\delta}^{\R\R\R\R} = \cWIR_{1234,\delta}^{\R\R\R\R} + \cWII_{1234,\delta}^{\R\R\R\R}, 
\end{align}
with 
\begin{align}
 \cWIR_{1234,\delta}^{\R\R\R\R} &= \CAcub \sum_{(ij) \in \Delta_\delta} \cC_{ij} \Big[ 
 \cA_{ij}^{12} \cAb_{ij}^{13} \cAb_{ij}^{14} + \cC_{ij}^{1234} +
\notag\\
& +  \sum_{sk\ell=2}^4 \cA_{ij}^{1s} \cBb_{ij}^{1k\ell} + \fA_{ij}^{1234} 
+ \frac{\cQ_\delta}{\CAcub} \bA_{ij}^{1234}
 \Big],
 \label{eq:EikAmp-4loop-IR}
 \\
 \cWII_{1234, \delta}^{\R\R\R\R} &= \CA\, \cQ_\delta\; \cNb_{1234}^{\R\R\R\R},
 \label{eq:EikAmp-4loop-II}
\end{align}
\end{subequations}
where $\bA_{ij}^{k\ell m n} = w_{ij}^k \bAb_{ij}^{k \ell m n}$ and $\cNb_{1234}^{\R\R\R\R}$ is the new irreducible contribution at this order. Notice that the contribution \eqref{eq:EikAmp-4loop-IR} is reducible in the sense that it is related to contributions from previous orders. In particular, the appearance of the quadrupole colour coefficient $\cQ_\delta$ is associated with the three-loop (quadrupole) antenna function \eqref{eq:quadrupoleFun} (while $\cC_{ij}$ is associated with $w_{ij}^k$). Eqs. \eqref{eq:EikAmp-4loop-R} and \eqref{eq:EikAmp-4loop-IR} hint at the {\it exponential} nature of eikonal amplitudes squared, as reported in \cite{Delenda:2015tbo}.  

The other antenna functions in \eqref{eq:EikAmp-4loop-IR} are defined by:
\begin{subequations}
\begin{align}
 \cC_{ij}^{k\ell m n } &= w_{ij}^k \left( \cB_{ik}^{\ell m n} + \cB_{kj}^{\ell m n } - \cB_{ij}^{\ell m n}  \right), 
 \\
 \fA_{ij}^{k\ell m n } &= w_{ij}^k \left( \cA_{ik}^{\ell m} \cAb_{ik}^{\ell n} + \cA_{kj}^{\ell m} \cAb_{kj}^{\ell n} - \cA_{ij}^{\ell m} \cAb_{ij}^{\ell n} \right).
\end{align}
\end{subequations}
There are 4 out of the 8 amplitudes squared mentioned above that do contain new (quadrupole) contributions. These are, in addition to RRRR: RRVR, RVRR and RVVR. Every single one of the said 4 amplitudes squared breaks both Bose and mirror symmetries \footnote{As discussed in details in Ref. \cite{Delenda:2015tbo}.}. We note that they add up to zero. That is: 
\begin{align}\label{eq:Sum-N-funs}
 \cNb^{\R\R\R\R}_{1234} +  \cNb^{\R\V\R\R}_{1234} +  \cNb^{\R\R\V\R}_{1234} +  \cNb^{\R\V\V\R}_{1234} = 0. 
\end{align}
The explicit expressions of the four $\cNb$ functions are cumbersome and have not been successfully written in a closed compact form. We provide the latter expressions in the accompanying \texttt{Mathematica} notebook file ``\texttt{3HL-N.nb}".  

It is worth mentioning that the \texttt{EikAmp} program can compute higher-loop squared amplitudes through the implementation of Eq. \eqref{eq:EikAmp_2HardLegs_Final} (and the relevant modifications for three-hard legs). Their expressions are, however, quite tedious and thus we do not consider them explicitly here in this letter. 
 
\section{Conclusion}
\label{sec:Conclusion}

The current work is an extension of our previous publication \cite{Delenda:2015tbo}. We have presented a general formula for the computation of QCD eikonal amplitudes for the emission of soft energy-ordered gluons off arbitrary two and three massless hard legs. The \texttt{Mathematica} program of \cite{Delenda:2015tbo} has been used to provide an automated way for computing all required eikonal squared amplitudes at a given order in PT (after implementing the appropriate modifications for the three-hard legs case).  

We have provided explicit expressions for the eikonal amplitudes squared for the various channels and gluon configurations for the three-hard legs case up to four loops in PT. We have shown that new ``quadrupole" functions first appear at three-loops which cannot be related to ``dipole" functions at previous orders. These quadrupole contributions have some peculiar features and were discussed in detail in \cite{Delenda:2015tbo}. 
We find agreement with available results \cite{Catani_2020} up to three-loops.
To the best of our knowledge the four-loops eikonal amplitude squared has not been previously reported in literature and is presented here for the first time. As was mentioned in the introduction, the general form of the eikonal amplitudes squared and the explicit expressions reported here are valid for any three massless hard legs QCD processes including, for instance, V/H+jet processes at hadron colliders and dijet production in DIS.

The treatment of eikonal amplitudes for the case of four-hard legs is quite delicate as the partonic Born processes involve more than one channel which are colour independent. This makes the calculations matrix-valued and thus more involved. We will postpone this work to our future publications.   
It is worth mentioning that the results of the current letter are employed in \cite{Khelifa-Kerfa} in the computation of the fixed-order distribution of QCD non-global observables for V/H+jet processes at hadron colliders up to fourth order in PT.

\section*{Acknowledgements}

We would like to thank
\begin{itemize}
	\item The Deanship of Research at the Islamic University of Madinah (research project ``Tamayyuz" No. 02/40), and 
	\item PRFU: B00L02UN050120190001 (Algeria).
\end{itemize}
for supporting this project. 

\bibliography{Refs}

\end{document}